# Numerical investigation of spallation neutrons generated from petawatt-scale laser-driven proton beams

B. Martinez[1,2*], S. N. Chen[3], S. Bolaños[1], N. Blanchot[4], G. Boutoux[2], W. Cayzac[2], C. Courtois[2], X. Davoine[2,5], A. Duval[2], V. Horny[1,2], I. Lantuejoul[2], L. Le Deroff[4], P. E. Masson-Laborde[2,5], G. Sary[2,5], B. Vauzour[2], R. Smets[6], L. Gremillet[2,5, †], J. Fuchs[1, §]

1. LULI-CNRS, CEA, UPMC Univ Paris 06: Sorbonne Université, Ecole Polytechnique, Institut Polytechnique de Paris – F-91128 Palaiseau Cedex, France
2. CEA, DAM, DIF, F-91297 Arpajon, France
3. Horia Hulubei National Institute of Physics and Nuclear Engineering, Bucharest – Magurele, Romania
4. CEA, DAM, CESTA, F-33114 Le Barp, France
5. Université Paris-Saclay, CEA, LMCE, 91680 Bruyères-le-Châtel, France
6. LPP, Sorbonne Université, CNRS, Ecole Polytechnique, F-91128 Palaiseau, France

* bertrand.martinez@tecnico.ulisboa.pt; Present address: GoLP/Instituto de Plasmas e Fusão Nuclear, Instituto Superior Técnico, Universidade de Lisboa, Lisbon, Portugal
† laurent.gremillet@cea.fr
§ julien.fuchs@polytechnique.edu

**Abstract**

Due to their high cost of acquisition and operation, there are still a limited number of high-yield, high-flux neutron source facilities worldwide. In this context, laser-driven neutron sources offer a promising, cheaper alternative to those based on large-scale accelerators, with, in addition, the potential of generating compact neutron beams of high brightness and ultra-short duration. In particular, the predicted capability of next-generation petawatt (PW)-class lasers to accelerate protons beyond the 100 MeV range should unlock efficient neutron generation through spallation reactions. In this paper, this scenario is investigated numerically through particle-in-cell and Monte Carlo simulations, modeling, respectively, the laser acceleration of protons from thin-foil targets and their subsequent conversion into neutrons in secondary heavy-ion targets. Laser parameters relevant to the 1 PW LMJ-PETAL and 1-10 PW Apollon systems are considered. Under such conditions, neutron fluxes exceeding $10^{23}$ n cm$^{-2}$s$^{-1}$ are predicted, opening up attractive fundamental and applicative prospects.

## 1. Introduction

Neutron beams are commonly employed in research, medicine and industry for a wide range of applications [1]. In practice, they are generated from nuclear reactions initiated by accelerator proton beams. Conventional neutron source facilities range from compact tubes to large-scale linacs like the Spallation Neutron Source (Oak Ridge, USA) [2] or the European Spallation Source (Lund, Sweden) [3] currently under construction, where $1 - 2$ GeV protons are hitting a heavy metal target to produce neutrons through spallation reactions. These consist of a cascade of binary collisions between the incident projectile and the nucleons inside the target nuclei, followed by de-excitation (or *evaporation*) of the excited nuclei, leading to the emission of neutrons, but also, to a smaller extent, protons, alpha particles, light heavy ions, gamma rays, etc. [4].

The production of bright neutron beams using high-power, short-pulse lasers was demonstrated in the early 2000's (see Ref. [5] and references therein for an overview) and has since been actively investigated. Laser-generated neutrons have already been utilized for a variety of purposes, such as material testing for fusion experiments [6], non-destructive imaging [7], and studies of equations of state via neutron resonance spectroscopy [8,9]. Such neutron sources exploit either laser-driven energetic protons (with current record-high energies of ~ 100 MeV [10]) [11], electrons [12] or gamma-ray photons [13] as the primary driver, with typical cross sections in the barn range [5,14] Most previous studies on this topic were focused on improving the yield [12,14] and the energy spectrum [15] of the emitted neutrons.

From a fundamental physics point of view, high-brightness neutron sources would be highly desirable for investigations into the formation of heavy elements in the Universe [16,17]. Half of the elements heavier than Iron ($Z = 26$), and all those beyond Bismuth ($Z = 83$), are indeed believed to originate from rapid neutron captures (r-process) during cataclysmic astrophysical events, e.g., supernova explosions or neutron star mergers [18]. In order to compensate for the short lifespan of the intermediate isotopes, a minimum neutron flux $> 10^{20}$ n cm$^{-2}$ s$^{-1}$ is estimated for the r-process to operate [19]. This value is several orders of magnitude above the capability of conventional accelerator-based facilities (~$10^{15}$ n cm$^{-2}$s$^{-1}$) [20], but also significantly larger than the current record-high flux (~$10^{18}$ n cm$^{-2}$s$^{-1}$) obtained with intense short-pulse lasers [12,14]. Neutron fluxes as high as ~$10^{24}$ n cm$^{-2}$s$^{-1}$ can be attained at large-scale laser fusion facilities [21], yet with limited user access and very few shots per experiment. Systematic laboratory investigations of *r-process* nucleosynthesis therefore require laser-based neutron sources to be further developed.

Laser acceleration of proton beams should greatly benefit from next-generation petawatt (PW) or multi-petawatt facilities, delivering pulse intensities in excess of $10^{21}$ Wcm$^{-2}$ [22,23,24,25,26,27,28,29]. At such intensities, the dominant ion acceleration mechanism is expected to transition from target normal sheath acceleration (TNSA) [30] to radiative pressure acceleration (RPA) or light-sail acceleration (LSA) [31]. The accompanying increase in proton energy above the 100 MeV level should trigger spallation reactions in a secondary neutron-producing target, entailing the emission of multiple neutrons per incident proton (see Figure 1).

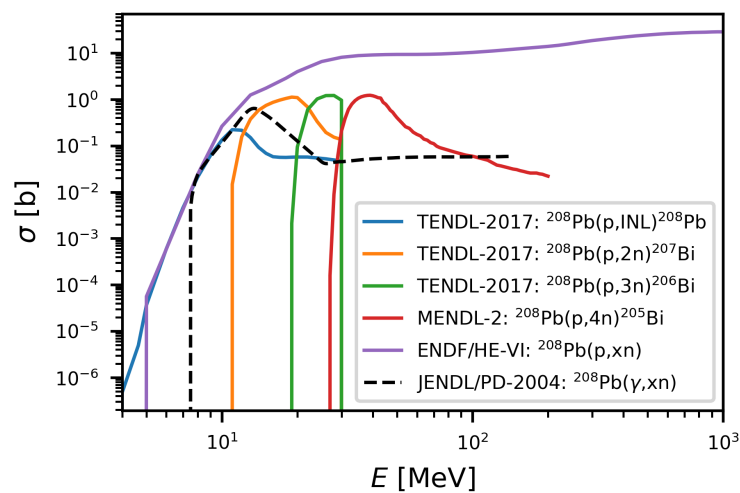

*Figure 1: Energy-differential cross sections of proton-induced nuclear reactions releasing different numbers of neutrons (solid curves) and of total neutron production by photonuclear reactions (black dashed curve) in Pb, as given by the ENDF/B-VIII database [32].*

In Figure 1, the cross sections of various neutron-producing reactions (as given by the ENDF/B-VIII database [32]) are plotted as a function of the projectile (proton or photon) energy (solid lines). Already at relatively low proton energies of tens of MeV, easily produced by current high-intensity laser facilities, neutrons are efficiently produced. Then, to access the reactions with increased neutron multiplicity, it is clear that higher projectile energies are needed.

It should be noted that the photonuclear reaction of lead (black dashed curve) has a comparable cross section around 10 MeV to that of proton-induced reactions, but is 100 times lower at higher energy. Hence, in the present study, we will focus on exploiting protons as a driver to induce the desired neutron beam. Photonuclear reactions will therefore be neglected in a first-order approach, but could prove to be influential at more extreme laser intensities ($\geq 10^{23}$ W.cm$^{-2}$, not presently achievable) for which massive high-energy synchrotron radiation is expected to arise [33].

This paper reports on a numerical study aiming to characterize the neutron yield and flux achievable at the petawatt-class LMJ-PETAL [34] and Apollon [35] laser facilities. Our results are also relevant to similar laser systems, such as ELI-beamlines [36] or ELI-NP [37]. Overall, this is done by combining particle-in-cell (PIC) simulations of the laser-driven particle acceleration in a thin-foil target with Monte Carlo calculations of the proton-induced nuclear reactions in a secondary target. In Section 2, we first investigate numerically how the production of spallation neutrons varies with the target material and incident proton energy. In Section 3, we present the results of PIC simulations of laser-based proton acceleration from solid foils under conditions accessible at LMJ-PETAL and Apollon. The generation of spallation neutrons from a lead convertor by the PIC-predicted proton beams is studied via Monte Carlo simulations in Sec. 4, which predict that neutron yields of $10^8 - 10^{11}$ n str$^{-1}$ and neutron fluxes of $10^{23} - 10^{24}$ n cm$^{-2}$ s$^{-1}$ are achievable. Finally, our results are summarized and discussed in Sec. 5.

## 2. Dependence of neutron production on target material and proton energy

| Proton energy (MeV) | Al | Cu | Ag | Pb |
|---|---|---|---|---|
| 25 | 0.315 | 0.117 | 0.115 | 0.135 |
| 50 | 1.08 | 0.391 | 0.380 | 0.435 |
| 100 | 3.70 | 1.31 | 1.26 | 1.43 |
| 250 | 17.9 | 6.28 | 5.97 | 6.64 |
| 500 | 55.0 | 19.1 | 18.1 | 19.9 |
| 1000 | 152 | 52.9 | 49.7 | 54.2 |

*Table 1: Projected range λ (cm) for protons in various materials and for various energies.*

We start by briefly examine the neutron yield from spallation as a function of the converter target material and the projectile proton energy in a range within the reach of present or near-future laser systems [10,31]. For this purpose, we have used the FLUKA 3D Monte Carlo code [38,39] to simulate the nuclear reactions induced during irradiation of a convertor target by a mono-kinetic and mono-directional proton beam. The proton beam energy was varied from $\epsilon_p = 25$ MeV to 1 GeV. Four target materials were considered:

aluminum ($Z = 13$), copper ($Z = 29$), silver ($Z = 47$), and lead ($Z = 82$). The rationale for considering materials lighter than Pb (i.e., a standard material for spallation purposes) is that the $\sim 100$ MeV maximum energies nowadays attained by laser-accelerated protons [10, 40] are still way lower than the $\sim$ GeV proton energies involved at conventional spallation neutron facilities. Hence, a high-$Z$ convertor target could incur collisional stopping of the incident protons before spallation takes place. Moreover, the less energetic neutrons produced by laser-driven protons will be more susceptible to collisional slowing down and scattering.

The simulated target was a 50 cm-radius cylinder of variable length $L$. Introducing $\lambda$ the energy-dependent projected range of a proton due to ionization and excitation [41], five $L/\lambda$ values (0.2, 0.4, 0.6, 0.8, 1) were considered for each material and input proton energy. The values of $\lambda$ corresponding to our parameter range are given in Table 1.

Figure 2 shows the $L/\lambda$ dependence of the neutron multiplicity $M_n$, i.e., the number of neutrons produced per incident proton as a function of its input energy $\epsilon_p$, for the different materials under consideration. The main result is that, whatever the material, $M_n$ rises sharply (i.e. approximately quadratically) with $\epsilon_p$ when $L/\lambda$ is kept constant. In Ag and Pb, $M_n$ approaches unity (a usual criterion for the onset of spallation) for $\epsilon_p \simeq 250$ MeV and $L/\lambda \lesssim 1$. At higher $\epsilon_p$, $M_n \geq 1$ can be achieved in lower $L/\lambda$ targets. The maximum neutron multiplicity ($M_n \sim 10$) is obtained in Pb with $\epsilon_p = 1$ GeV and $L/\lambda = 0.4 - 0.6$. It should be noted that the increasing trend of $M_n$ with $\epsilon_p$ ceases beyond $\epsilon_p \simeq 0.5 - 1$ GeV when $L/\lambda \gtrsim 0.5$. This is a known behavior in spallation studies, ascribed to the increasingly significant contribution of pion production to proton energy losses [4].

At fixed proton energy, $M_n$ is predicted to rise by a relatively modest ($\sim 3\times$) factor when the normalized target thickness is increased from $L/\lambda = 0.2$ to 1. Finally, at fixed $\epsilon_p \leq 0.5$ GeV and $L/\lambda$, $M_n$ shows a moderate increase with the atomic number, that is, a $\sim 3\times$ enhancement between Al and Pb. At $\epsilon_p = 1$ GeV, however, an approximate $10\times$ enhancement is obtained.

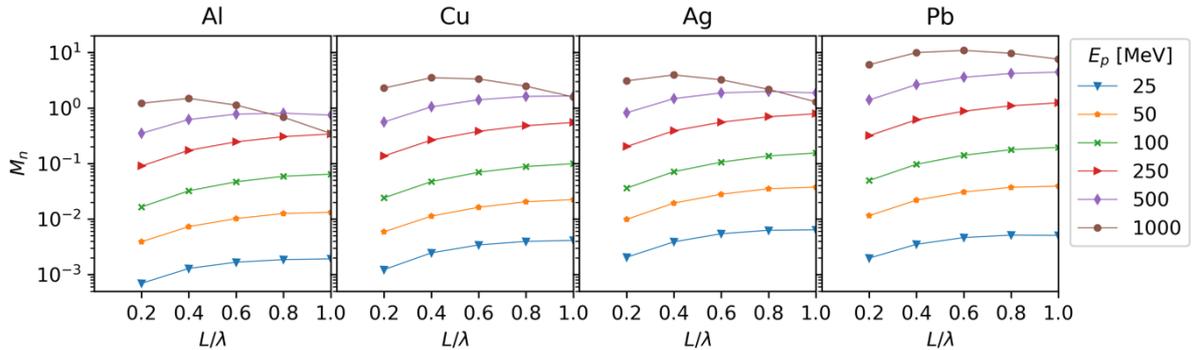

Figure 2: Number of neutrons emitted per incident proton as a function of the target material and incident proton energy, as simulated by FLUKA.

## 3. PIC simulations of laser acceleration of protons

We now numerically characterize the neutron beams that could be produced through spallation reactions in the near future, using the LMJ-PETAL and Apollon laser systems. Our methodology comprises two steps. First, we have employed the CALDER code [42] to perform multidimensional PIC simulations of proton acceleration from laser-irradiated foil targets, under conditions relevant to the LMJ-PETAL and Apollon lasers, as detailed in

Table 2. Second, the proton distributions recorded in the PIC simulations have been used as input in 3D FLUKA Monte Carlo simulations [39], describing the proton transport and associated nuclear reactions through a secondary Pb convertor target. This procedure is sketched in Figure 3. In this Section, we present the results of the proton acceleration simulations for the LMJ-PETAL and Apollon cases.

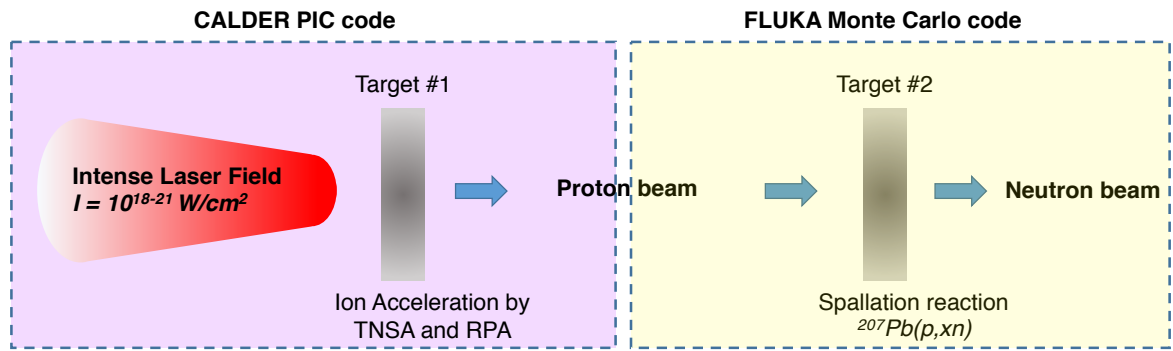

Figure 3: *Conceptual setup of the numerical study.*

| Laser | Wavelength | Pulse duration | Pulse energy | Pulse intensity | Target size and composition | Simulation mesh size |
|---|---|---|---|---|---|---|
| 1 PW LMJ-PETAL | 1 μm | 610 fs | 320 J | $2 \times 10^{18}$ W/cm$^2$ | 5 μm CH & Al | 32 nm |
| 1 PW Apollon | 0.8 μm | 20 fs | 12 J | $2 \times 10^{21}$ W/cm$^2$ | 64 nm CH | 3.2 nm |
| 10 PW Apollon | 0.8 μm | 20 fs | 120 J | $2 \times 10^{22}$ W/cm$^2$ | 192 nm CH | 3.2 nm |

Table 2: *Parameters of the 2D CALDER PIC simulations performed for each considered laser system.*

### 3.1 Proton acceleration at the 1 PW LMJ-PETAL laser facility

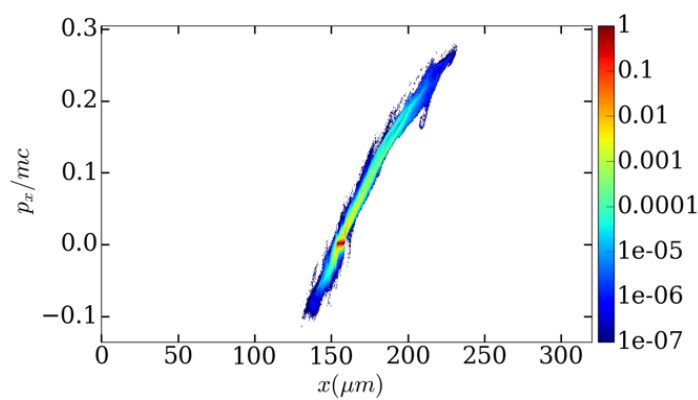

Figure 4: *Longitudinal $(x - p_x)$ phase space of the protons from the CALDER-CIRC simulation using the LMJ-PETAL parameters.*

Proton acceleration using LMJ-PETAL was investigated in quasi-3D geometry with the CALDER-CIRC PIC code [43]. Based on experimental measurements [44], the PETAL laser beam was modeled as two superimposed, time-synchronized Gaussian laser waves. Both propagated along the longitudinal $x$-axis, were linearly polarized along the $y$-axis, and had a 610 fs FWHM duration. The first, corresponding to the central region of the laser pulse, had a FWHM width of $20\,\mu m$ and a dimensionless field strength of $a_L = eE_L/m_e c\omega_L = 1.1$ (associated with a $\sim 1.7\times10^{18}\,\text{Wcm}^{-2}$ peak intensity); the second, representing the wings of the focal spot, had a FWHM width of $130\,\mu m$ and a dimensionless field strength of $a_L = 1.3$ (associated with a $\sim 2.3\times10^{18}\,\text{Wcm}^{-2}$ peak intensity). The cumulated maximum laser intensity was of $8\times10^{18}\,\text{Wcm}^{-2}$.

The target consisted of a semi-transparent aluminum (Al) preplasma and an overcritical plastic (CH) plasma (5 $\mu$m thick). Henceforth, $n_c = m_e\epsilon_0\omega_L^2/e^2$ ($\omega_L$ is the laser frequency, $m_e$ the electron mass, $e$ the elementary mass and $\epsilon_0$ the vacuum permittivity) will denote the critical density beyond which the laser can no longer propagate in the plasma. For the $\lambda_L = 1\,\mu m$ wavelength of the PETAL laser pulse, one has $n_c = 1.1\times10^{21}\text{cm}^{-3}$. Based on hydrodynamic-radiative simulations [45], the electron density profile of the preplasma was taken to evolve as $n_e(x) = 5\,n_c\exp(5.45(x/150)^{0.29})$, where the longitudinal position $x$ is here expressed in $\mu$m units. The minimum and maximum density values were set to $0.02\,n_c$ and $5\,n_c$. The latter maximum density, also characterizing the uniform CH layer, was chosen high enough to describe accurately the absorption of the laser light and low enough to relax the constraints on the numerical discretization. The Al, C and H ions were assumed fully ionized, and initialized at a 10 eV temperature. The ionic species $Al^{13+}$, $C^{6+}$, $H^+$ and the electrons were represented with 4, 4, 8 and 32 macro-particles per cell, respectively. Coulomb collisions between the plasma particles were neglected. The laser wave was $p$-polarized and impinged normally onto the plasma. Absorbing boundary conditions were used for both particles and fields. The simulation domain, of $L_x \times L_r = 318\,\mu m \times 163\,\mu m$ dimensions, was discretized with a $\Delta x = \Delta r = 32$ nm mesh size. The simulation was run during a $\sim 3.4$ ps integration time.

With the above parameters, proton acceleration proceeds from the standard TNSA mechanism [15]. During the interaction, the laser wave propagates through the extended undercritical preplasma while driving the electrons to relativistic energies through various processes. These have been examined in detail by some of us in a recent study [44], which revealed the importance of stochastic electron heating as a result of stimulated laser backscatter and laser filamentation in the preplasma.

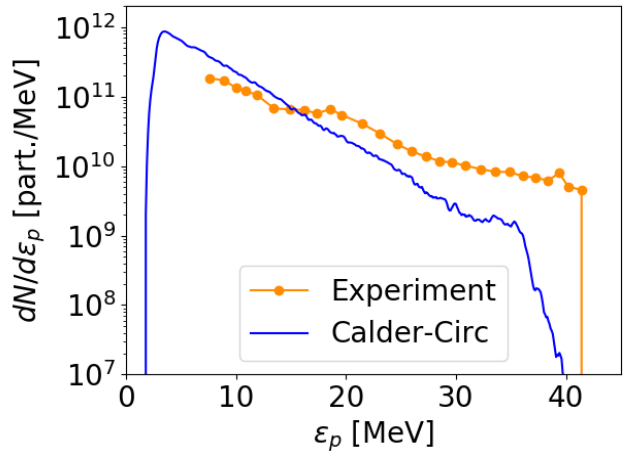

*Figure 5: Proton spectrum from the CALDER-CIRC simulation using the LMJ-PETAL laser parameters (blue curve). An experimental proton spectrum obtained at LMJ-PETAL (see text for details) is plotted as orange dots.*

After traversing the target, the laser-generated hot electrons form a negatively charged cloud at the backside. The associated electrostatic field then accelerates the plasma ions in the $+x$ direction. Figure 4 displays the longitudinal phase space of the protons as measured at $t = 3.4$ ps. The proton distribution exhibits a linear shape typical of TNSA, extending over $\sim 100\ \mu m$ longitudinally and (from analysis of the $x - r$ proton density map) $\sim 300\ \mu m$ transversely. The corresponding energy spectrum of the protons is plotted in Figure 5. It shows a decreasing exponential shape characteristic of TNSA, with a cutoff energy of $\sim 37$ MeV.

To support the validity of our approach, we compare in Figure 5 the proton distribution predicted by CALDER-CIRC to an experimental one, which was recorded by us at LMJ-PETAL. It was measured with the CRACC diagnostic, which uses a radiochromic film stack as detector; the spectrum is here shown after angular integration onto the whole surface of the films [46]. In this shot, the PETAL laser irradiated a $7\ \mu m$ thick titanium foil with a pulse of $960\ fs$ duration, $354\ J$ energy and $\sim 5.3 \times 10^{18}\ Wcm^{-2}$ intensity, i.e., parameters relatively close to those of the simulation. Although the experimental proton spectrum is characterized by a harder slope and a larger cutoff energy ($\sim 42\ MeV$) than the simulated spectrum, the two curves agree fairly well with each other. This result is particularly satisfactory given the above simplifications made to the simulation in order to handle the spatiotemporal scales of the experiment. It should be noted that the experimental cut-off energy represents the last recorded point on the RCF stack although there are more RCFs in the stack); such sharp cutoff is a common feature of TNSA-accelerated proton beams

## 3.2 Proton acceleration at the 1-10 PW Apollon laser facility

Proton acceleration using Apollon was simulated in 2D Cartesian geometry with the CALDER PIC code. The Apollon pulse was modeled as a $0.8\ \mu m$ wavelength Gaussian electromagnetic wave with 20 fs FWHM duration and $5\ \mu m$ FWHM transverse size. We considered two operating regimes of the Apollon laser, characterized by a peak intensity of $2 \times 10^{21}\ Wcm^{-2}$ ("1 PW regime") and $2 \times 10^{22}\ Wcm^{-2}$ ("10 PW regime"), respectively.

The target was a thin, solid-density CH foil with sharp gradients. It was assumed to be fully ionized (yielding an electron density of $n_e = 200 n_c$) and initialized at a temperature of 100 eV. Its thickness $l$ was chosen based on the parametric simulation study of Ref. [37]. The optimum foil thickness for RPA by femtosecond laser pulses was found to be $l_{opt} \simeq 0.5 a_L (n_c/n_e) \lambda_L$. In the 1 PW regime ($a_L = 30$), we chose $l = 64$ nm, close to $l_{opt} = 60$ nm. Each of the plasma constituents ($C^{6+}, H^+$, electrons) was modeled by 100 macro-particles per cell. The laser pulse was *p*-polarized and interacted at normal incidence with the target. Absorbing boundary conditions were enabled for the fields and particles. The simulation domain was set to $L_x \times L_y = 40 \times 50\ \mu m^2$ at 1 PW ($56 \times 96\ \mu m^2$ at 10 PW) with a spatial resolution $\Delta x = \Delta y = 3.2$ nm.

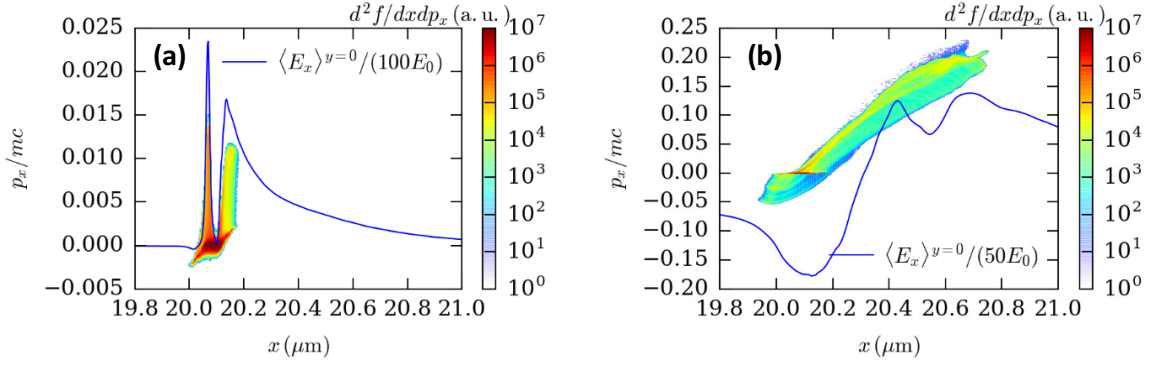

*Figure 6: Proton acceleration using the 1 PW Apollon laser parameters: $x - p_x$ proton phase spaces at (a) $t = -20\,fs$ and (b) $t = +4\,fs$ (here $t = 0$ corresponds to the on-target laser pulse maximum). The blue line is the laser-cycle-averaged longitudinal electric field, $\langle E_x \rangle$, normalized to (a) $100\,E_0$ or (b) $50 E_0$ for readability ($E_0 = 3.2 \times 10^{12}\,Vm^{-1}$).*

The proton acceleration dynamics is illustrated by the longitudinal proton phase spaces shown at two successive times in Figure 6 and Figure 7 for the 1 PW and 10 PW irradiation cases, respectively. In each panel is also plotted (in blue) a lineout (along the laser axis) of the accelerating longitudinal electric field $\langle E_x \rangle$ (here $\langle \; \rangle$ indicates an average over the laser cycle). This field is normalized (in units of $E_0 = 3.2 \times 10^{12}\,Vm^{-1}$) to fit within the $p_x$-axis of the phase space.

At time $t = -20$ fs (here the time origin $t = 0$ is when the laser maximum reaches the target), see panels (a) in Figure 6 and Figure 7, proton acceleration originates from both RPA and TNSA [9, 29], as demonstrated by the two $\langle E_x \rangle$ peaks at the front and rear sides of the target. At the front side, the electrons are pushed and compressed by the laser's ponderomotive force. The ensuing charge separation generates an electrostatic field $\langle E_x \rangle \simeq 2 E_0$ at 1 PW ($\simeq 10 E_0$ at 10 PW), which, in turn, accelerates the front-side protons in the forward direction. These RPA protons have then reached a longitudinal momentum $p_x/m_i c \simeq 0.015$. Simultaneously, the backside protons have started expanding towards vacuum due to TNSA triggered by the fast electrons. The associated electric field $\langle E_x \rangle \simeq 1.5 E_0$ ($\simeq 3 E_0$ at 10 PW), however, turns out to be weaker than the one induced by the radiation pressure (especially at 10 PW). At this early stage of the interaction, the maximum momentum of the TNSA protons is of $p_x/m_i c \simeq 0.0011$ ($\simeq 0.02$ at 10 PW).

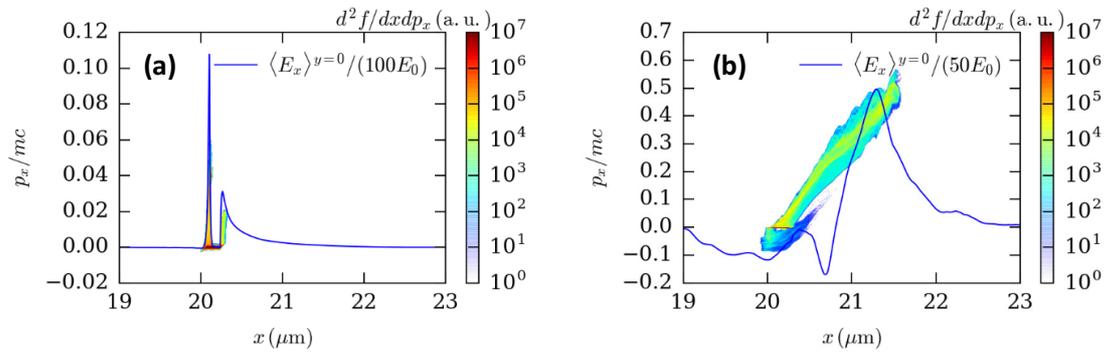

*Figure 7: Proton acceleration using the 10 PW Apollon laser parameters: $x - p_x$ proton phase spaces at (a) $t = -20\,fs$ and (b) $t = +4\,fs$ (here $t = 0$ corresponds to the on-target*

*laser pulse maximum). The blue line is the laser-cycle-averaged longitudinal electric field, $\langle E_x \rangle$, normalized to (a) $100\,E_0$ or (b) $50 E_0$ for readability ($E_0 = 3.2 \times 10^{12} Vm^{-1}$).*

At $t = +4$ fs [panels (b) in Figure 6 and Figure 7], the RPA protons have caught up with the TNSA protons, an the two previously observed field structures have merged into a single accelerating structure. Importantly, the expanding, lower-density plasma has then turned transparent to the central part of the laser pulse. This causes the electrons to be further heated and the accelerator field strength to be boosted to $\langle E_x \rangle \simeq 7.5\,E_0$ at 1PW and $\langle E_x \rangle \simeq 25\,E_0$ at 10 PW), which, in both cases, represents about a quarter of the laser field amplitude.

As a result of this sequence of processes, the protons eventually attain high cutoff energies ($\simeq 115$ MeV at 1 PW and $\simeq 660$ MeV at 10 PW) as demonstrated by the energy spectra displayed in Figure 8(a,b). Note that these spectra are recorded when the protons reach the right-hand side of the box: their spatial distribution then has a $\sim 20 - 30\,\mu m$ transverse size, comparable with the travelled distance. The electrostatic field seen by the fastest protons should therefore be relatively well captured by our 2D simulation. We acknowledge, however, that the combination of a 2D geometry and $p$-polarized laser pulse likely leads to a significant overestimation of the electron heating, and hence of the accelerator field compared to a real-world 3D configuration [47].

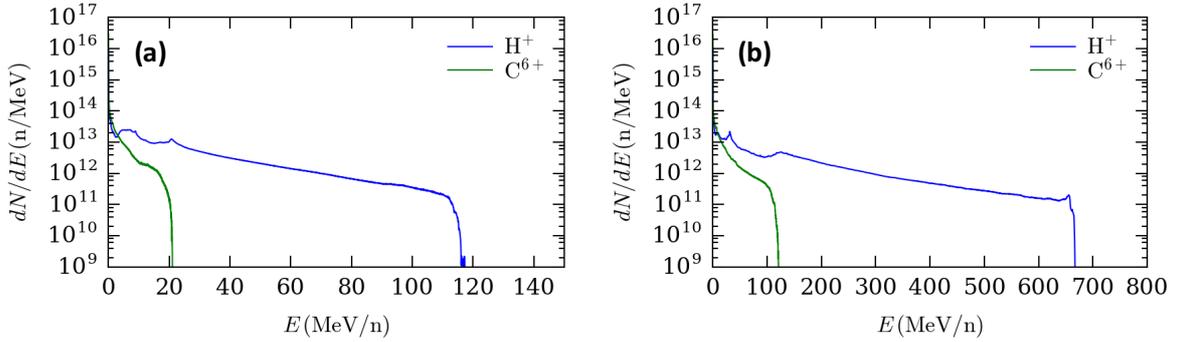

*Figure 8: Proton spectra from the CALDER code using the (a) 1 PW and (b) 10 PW Apollon laser parameters.*

### 4. Monte Carlo simulations of neutron generation

In the CALDER simulations, the properties (statistical weight, position, momentum, time of arrival) of the macro-protons crossing a virtual detector plane near the right-hand side of the simulation box were all recorded. The resulting output files contained about $10^5 - 10^6$ macro-protons. The proton distributions obtained from 2D PIC simulations had to be post-processed in order to be used as input in the FLUKA 3D Monte Carlo code. To this purpose, they were converted into cylindrically symmetric distributions. Specifically, the position and momentum of each macro-proton were rotated around the $x$-axis by a random azimuthal angle. Moreover, the transverse density profile of the proton distribution was interpreted as a radial density profile: the statistical weight of each macro-proton (a linear density in a 2D simulation) was thus multiplied by its transverse radius to obtain a dimensionless quantity, corresponding to the number of physical protons represented by the macro-proton.

The convertor target was taken to be a lead cylinder of fixed 5-cm radius and varying length ($10\,\mu m \leq l \leq 10$ cm), located 0.5 cm behind the proton-generating target. To get good statistics on the simulated events, we carried out 1000 independent Monte Carlo simulations for each set of initial conditions. Only those neutrons crossing the target rear side were characterized.

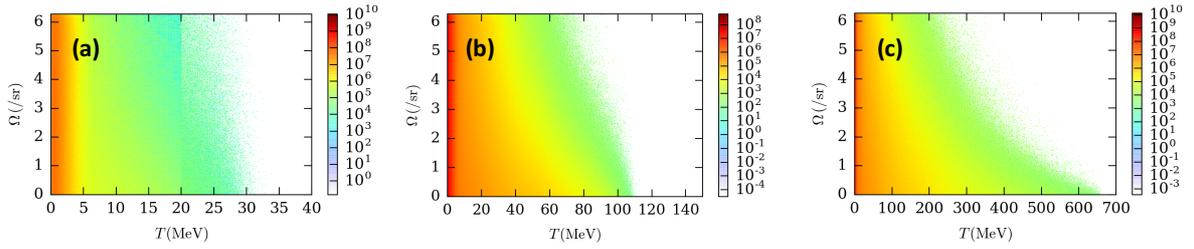

*Figure 9: Energy-angle spectrum of the neutrons escaping from the Pb convertor target. (a) The incident proton beam is that of the LMJ-PETAL PIC simulation. (b) Same with the 1 PW Apollon laser parameters (c) Same with the 10 PW Apollon laser parameters.*

Figure 9 displays the energy-angle spectrum of the outgoing neutrons as predicted by FLUKA in the LMJ-PETAL (a) and 1-10 PW Apollon (b,c) cases. Overall, the neutron energy spectra show an exponentially decreasing shape up to a maximum energy close to that of the incident protons. This is more clearly seen in Figure 12.b, obtained from angular integration of the spectra of Fig. 9. It is worth noting that the lower-energy part of the neutron distribution is essentially isotropic, while its higher-energy part is preferentially emitted in the initial direction of the proton beam.

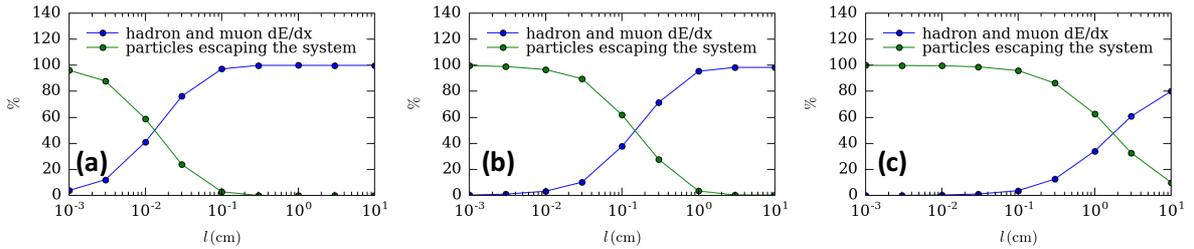

*Figure 10: Energy fraction of the incident protons dissipated by nuclear reactions (blue) and transmitted through the target (green), as a function of the thickness $l$ of the Pb converter target. Panels (a), (b) and (c) correspond, respectively, to the LMJ-PETAL, 1 PW Apollon and 10 PW and Apollon lasers.*

Figure 10 plots, as a function of the Pb target length $l$, the fraction of kinetic energy lost by the primary protons through nuclear reactions (blue curves) and that transmitted through the target (green curve). In the PETAL case, the proton beam energy is wholly dissipated for $l \geq 0.1$ cm. In the 1 PW Apollon case, this takes place for $l \geq 1$ cm, while in the 10 PW case, the absorption is limited to $\sim 80\%$ at $l = 10$ cm.

Figure 11 (a) shows how the total number (per unit solid angle) of outgoing neutrons varies with $l$. Under the LMJ-PETAL conditions, the proton number is seen to rise with $l$ up to $l \simeq 0.1$ cm and to saturate at a $\sim 10^8$ n sr$^{-1}$ level in the range $0.1 \leq l \leq 3$ cm, before dropping at larger $l$ as a result of reabsorption. In the 1 PW and 10 PW Apollon cases, due to higher proton energies, saturation occurs in thicker targets, namely, at $\sim 5 \times 10^8$ n sr$^{-1}$ for $1 \leq l \leq 3$ cm and at $\sim 10^{11}$ n sr$^{-1}$ $3 \leq l \leq 10$ cm, respectively (the plateau observed at 10 PW may actually extend beyond the range of thicknesses considered here). These trends are consistent with the evolution of the dissipated proton energy as discussed above. Interestingly, the neutron yield is predicted to be quite similar for the LMJ-PETAL and 1 PW Apollon lasers. At first glance, this result may seem surprising given that LMJ-PETAL generates about 50 times more fast protons than 1 PW Apollon ($\sim 5 \times 10^{12}$ in a $2 - 40$ MeV energy range vs. $\sim 10^{11}$ in a $10 - 120$ MeV range) due to its relatively large spot size and long pulse duration. Yet the lower proton number achieved at 1 PW Apollon is compensated for by their higher cutoff energy that increases the neutron generation efficiency through spallation. Such enhancement is even more dramatic using the 10 PW Apollon parameters, in which case a sharp rise in the proton energies is observed. The $\sim 5\times10^{11}$ protons then produced in a $20 - 700$ MeV energy range are predicted to translate into a two orders of magnitude higher neutron yield.

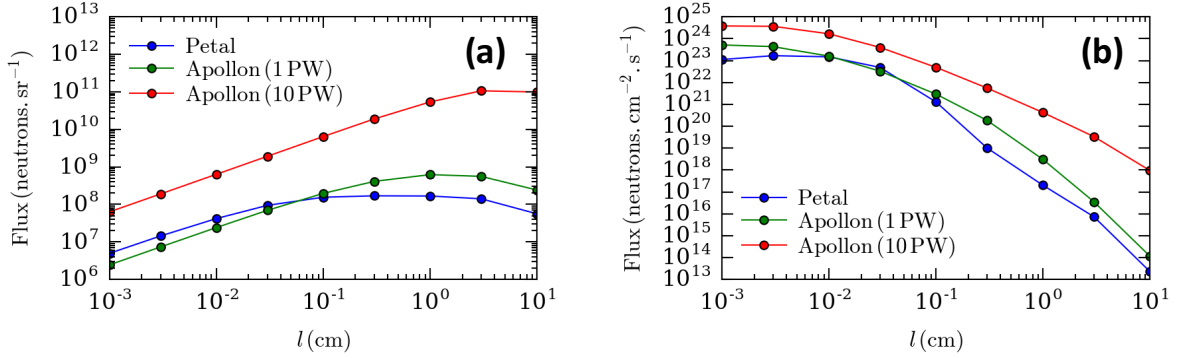

*Figure 11: (a) Number (normalized to unit solid angle) and (b) maximum flux of the neutrons crossing the rear side of the Pb converter target, as a function of its thickness l. The incident proton beam is that predicted by PIC simulations in the LMJ-PETAL and 1-10 PW Apollon cases, as labeled.*

Figure 11 (b) plots the corresponding variations in the maximum neutron flux at the backside of the Pb target. Note that FLUKA takes account of the time of injection of each proton into the convertor, so that the temporal profile of the total neutron flux across a given surface can be computed. The maximum flux appears to culminate in $l \lesssim 100$ $\mu$m targets and to drop in increasingly thick targets. Values in excess of $\sim 1\times10^{23}$n cm$^{-2}$ s$^{-1}$ are expected at LMJ-PETAL and 1 PW Apollon, while a maximum flux as high as $\sim 4\times10^{24}$n cm$^{-2}$s$^{-1}$ is found with the 10 PW Apollon parameters. This trend results from the increase in duration and the transverse size of the neutron distribution when the converter target is made thicker These variations originate from the energy dispersion of the incident proton beam (which leads to an elongation of the proton beam, and therefore of the generated neutron beam) as well as from elastic scatterings of both the protons and neutrons throughout the target (which mainly account for the transverse size of the neutron source). The temporal dependence of the neutron flux is illustrated in Figure 12(a) for LMJ-PETAL. It is observed that upon thickening the target from $l = 10$ $\mu$m to 1 cm, the duration of the neutron source rises from $\sim$ 3 ps to $\sim$ 6 ns. The correlation between the neutron source duration and transverse size is clearly shown in Figure 13. The three laser configurations give rise to a similar behavior: very compact neutron sources, of a few ps duration and $\sim 50 - 100$ $\mu$m width only, are expected from $l \leq 100$ $\mu$m Pb targets, which evolve into a few ns-long and cm-wide sources when cm-thick Pb targets are employed.

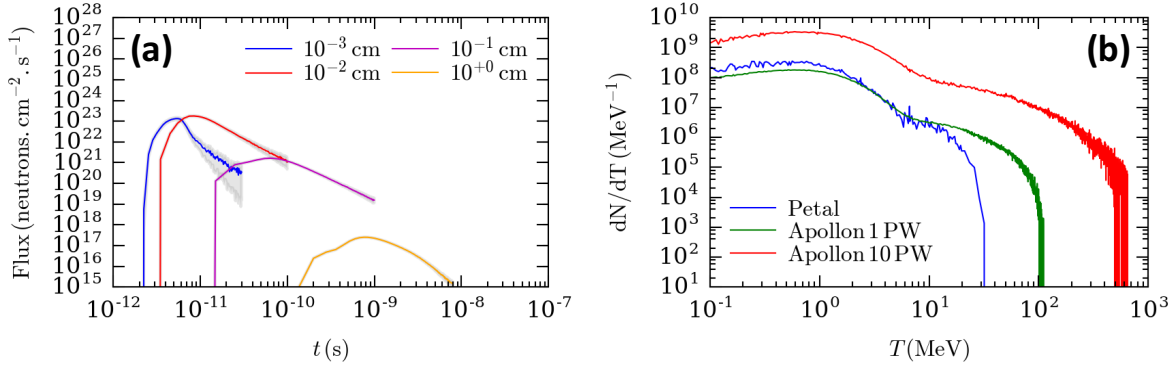

*Figure 12: (a) Time-dependent neutron flux across the Pb converter backside for the LMJ-PETAL parameters. (b) Neutron energy spectra from a l = 0.3 mm Pb target in the LMJ-PETAL and 1-10 PW Apollon cases.*

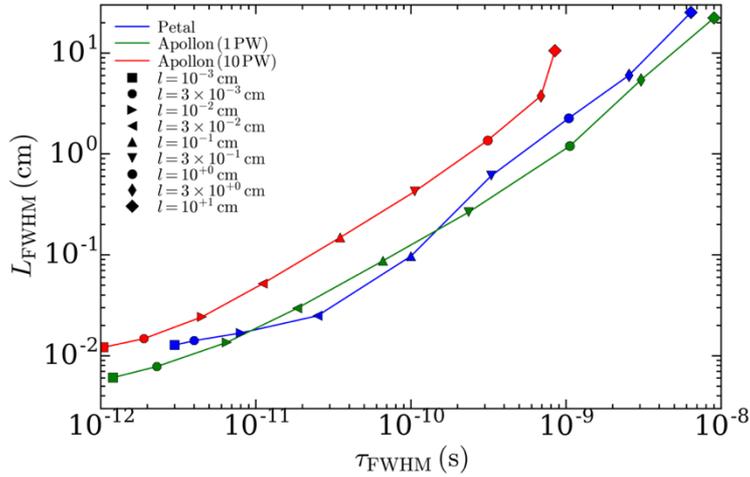

*Figure 13: Transverse size vs duration of the simulated neutron beam in the LMJ-PETAL, 1 PW Apollon and 10 PW Apollon cases, and for various thicknesses, as indicated.*

## 5. Discussion and summary of the properties of simulated neutron sources

We have here assessed the possibility of exploiting spallation reactions to generate high-flux neutron sources using PW-class lasers as the primary drivers. We have tested this scenario by combining two simulation codes, the first to simulate the proton generation by the laser, the second to simulate the neutron generation in a lead converter. Most notably, we have shown the interest of using ultraintense femtosecond laser pulses (here delivered by the 1 PW and 10 PW Apollon systems) to push the maximum proton energy beyond the 100 MeV threshold, that is, well above the level attained by higher-energy picosecond lasers such as the 1 PW LMJ-PETAL system. Such high proton energies translate into much larger neutron multiplicity from the convertor targets, and therefore allow the 1-10 PW laser systems to make up for their lower proton output.

Regarding the quantitative accuracy of our study, satisfactory agreement was demonstrated in the LMJ-PETAL case between an experimental proton spectrum and that obtained from a quasi-3D PIC simulation. While, as of now, such a comparison cannot be made in the Apollon setting, since the facility is still undergoing commissioning, we acknowledge the limitations of our 2D PIC simulations, and the fact that they may appreciably overestimate the proton cutoff energy (particularly in the 10 PW regime) as claimed by previous works [47]. This leaves room for a refined (but much more computationally demanding) simulation study based on 3D simulations to be conducted in the future.

To conclude, we note that the $> 10^{23}$ n cm$^{-2}$s$^{-1}$ peak neutron fluxes predicted by our numerical study are appropriate to laboratory studies of r-process nucleosynthesis. In particular, the short duration of the neutron source (shown in Figure 12.a and Figure 13), which varies from picosecond to ten nanoseconds, is adequate to perform nucleosynthesis experiments, since the β-decay of the created isotopes resulting from multiple neutron absorption occurs over much longer (> ms) timescales [48]. In short, the conditions for tackling r-process nucleosynthesis investigations in the laboratory seem good; with the start of the Apollon laser facility, and other facilities like ELI, one will soon be able to verify if the numerically predicted neutrons fluxes can be indeed achieved.

**Acknowledgements**

This work was supported by the European Research Council (ERC) under the European Union's Horizon 2020 research and innovation program (Grant Agreement No. 787539). It was also supported by Grant ANR-17-CE30-0026-Pinnacle from Agence Nationale de la Recherche. We acknowledge GENCI, France, for awarding us access to HPC resources at TGCC/CCRT (Allocation No. A0010506129). The PETAL laser was designed and constructed by CEA under the financial auspices of the Conseil Régional d'Aquitaine, the French Ministry of Research, and the European Union. The CRACC diagnostic was designed and commissioned on the LMJ-PETAL facility as a result of the PETAL+ project coordinated by University of Bordeaux and funded by the French Agence Nationale de la Recherche under grant ANR-10-EQPX-42-01. The LMJ-PETAL experiment presented in this article was supported by Association Lasers et Plasmas and by CEA. The diagnostics used in the experiment have been realized in the framework of the EquipEx PETAL+ via contract ANR-10-EQPX-0048.